\documentclass[a4paper,aps,prd,10pt,preprintnumbers,showpacs,twocolumn,superscriptaddress,nofootinbib,amsmath,amssymb]{revtex4-1}\usepackage{graphicx}\def\PRDstyle#1{#1}\def\JCAPstyle#1{}\def\Abstract#1{\begin{abstract}#1\end{abstract}}

\usepackage{cmap}
\usepackage[utf8]{inputenc}
\usepackage[T1]{fontenc}

\def\imo{i}
\def\a{\widetilde{\alpha}}

\newcommand{\ie}{{i.e.,}~}
\newcommand{\eg}{{e.g.,}~}

\begin{document}

\title{Overtones' outburst of asymptotically AdS black holes}

\JCAPstyle{
\author[1]{R.~A.~Konoplya,}\emailAdd{roman.konoplya@gmail.com}
\author[2]{A.~Zhidenko}\emailAdd{olexandr.zhydenko@ufabc.edu.br}
\affiliation[1]{Research Centre for Theoretical Physics and Astrophysics, \\ Institute of Physics, Silesian University in Opava, \\ Bezručovo náměstí 13, CZ-74601 Opava, Czech Republic}
\affiliation[2]{Centro de Matemática, Computação e Cognição (CMCC), \\ Universidade Federal do ABC (UFABC), \\ Rua Abolição, CEP: 09210-180, Santo André, SP, Brazil}
\arxivnumber{2310.19205}
}

\PRDstyle{
\author{R. A. Konoplya}\email{roman.konoplya@gmail.com}
\affiliation{Research Centre for Theoretical Physics and Astrophysics, Institute of Physics, Silesian University in Opava, Bezručovo náměstí 13, CZ-74601 Opava, Czech Republic}
\author{A. Zhidenko}\email{olexandr.zhydenko@ufabc.edu.br}
\affiliation{Centro de Matemática, Computação e Cognição (CMCC), Universidade Federal do ABC (UFABC), \\ Rua Abolição, CEP: 09210-180, Santo André, SP, Brazil}
}

\Abstract{Recently it was shown that small deformations of the asymptotically flat black-hole geometry in some region near its horizon which do not alter considerably the fundamental mode, nevertheless, strongly affect the first several (and higher) overtones which deviate at an increasing rate from their nondeformed limits. Here we show that, despite the quasinormal spectrum in anti-de Sitter (AdS) space is totally different from the asymptotically flat one, the outburst of overtones do take place at small near-horizon deformations of Schwarzschild-AdS spacetime as well.
Moreover, qualitatively new, nonoscillatory modes appear as a result of such small deformations, representing, thereby, a nonperturbative branch of modes. For this purpose we extend the general parametrization of asymptotically flat black holes to the AdS case. Small near-horizon deformations may originate from various holographically motivated factors, such as quantum corrections or attempts to describe the regime of intermediate coupling via higher curvature terms, while the nonoscillatory modes may be related to the hydrodynamic mode on the gauge theory side. Therefore, the phenomenon of the overtones' outburst must be taken into consideration when analyzing correlation functions and dispersion relations in the dual field theory.
In addition to the ad hoc deformations we consider the case of five-dimensional Einstein-Gauss-Bonnet-AdS black holes as an example of such near horizon deformations and fulfill the detailed study of the overtones.}

\maketitle

\section{Introduction}

In the AdS/CFT correspondence \cite{Maldacena:1997re,Gubser:1998bc,Witten:1998qj}, and, more generally, gauge-gravity duality, one of the fundamental insights is that quasinormal modes (QNMs) of black holes in the asymptotically anti-de Sitter (AdS) bulk correspond to poles of the retarded correlation functions in the dual quantum (conformal) field theory (CFT) \cite{Birmingham:2001pj,Son:2002sd,Kovtun:2005ev}. Thus, the quasinormal spectra directly determine the dispersion relations of excitations in the dual strongly coupled theory. Of particular interest are gapless QNMs of black holes with planar horizons: they correspond to hydrodynamic long-lived modes in the theory and therefore carry valuable information about its transport properties such as viscosity and conductivity \cite{Kovtun:2005ev,Son:2007vk}. This is relevant in the study of the quark-gluon plasma \cite{Romatschke:2007mq}, where strong coupling makes conventional perturbative methods ineffective.

Quasinormal spectrum of asymptotically AdS spacetimes \cite{Horowitz:1999jd,Cardoso:2003cj} is drastically different from those of flat or de Sitter spacetimes. The modes in AdS quickly achieve their high overtone number limit $n\to\infty$ being almost equally spaced in $n$. When the black-hole radius vanishes the modes approach the normal modes of the pure AdS spacetime \cite{Konoplya:2002zu}, so that one could interpret the spectrum as the AdS spectrum perturbed by the presence of a black hole. Unlike asymptotically de Sitter or flat case, the AdS modes form a full set and their superposition represents the signal at all times. This feature apparently follows from a different boundary conditions in AdS space, which serves as an effective confining box for perturbations.
These peculiarities of the AdS spectrum are worth mentioning, because the aspect we are interested here is closely related to our recent observations made in asymptotically flat spacetimes: small deformations of the black-hole geometry in the near-horizon zone will not affect the fundamental mode seemingly, but first several and higher overtones will deviate from their Schwarzschild limits at an increasing rate producing totally different picture of overtones \cite{Konoplya:2022pbc}. This observation is important mainly because of the two reasons: (a) The first several overtones are necessary to model the ringdown with a set of modes not only at its very end, but during the whole stage of damped quasinormal oscillations \cite{Giesler:2019uxc}. (b) The overtones are highly sensitive to the least deformations in the near event horizon zone, so they allow one to probe the near-horizon geometry, while the fundamental mode is dependent on the geometry near the peak of the potential barrier. Thus, the outburst of overtones represent a kind of ``the sound of the event horizon''. Consequently, an outburst of overtones has been considered in a number of black hole models which differ from the Schwarzschild spacetime by a relatively small deformation near the event horizon \cite{Konoplya:2023ahd,Bolokhov:2023ruj,Bolokhov:2023bwm,Konoplya:2023ppx,Konoplya:2023aph,Konoplya:2022iyn,Konoplya:2022hll}.

While overtones of asymptotically flat black holes are difficult to observe in the current and near future experiments, for the AdS case they have their own value within the AdS/CFT correspondence as poles of the retarded Green function in the dual field theory, expressing the dispersion relations on the finite temperature field theory side. At the same time, as the AdS spectrum is qualitatively different from the flat one, it is risky to state beforehand whether such outburst of overtones will take place in AdS or not.

Here we answer positively this question and find that the overtones not only explode by magnitude, but may change qualitatively as a result of such small near-horizon deformations, because the purely imaginary, \ie nonoscillatory, quasinormal modes appear in the spectrum. In order to ``prepare'' the near-horizon deformations we extend here the general parametrization of the spherical asymptotically flat black holes, proposed by L. Rezzolla and A. Zhidenko \cite{Rezzolla:2014mua} to the asymptotically AdS case in four and five dimensions.

We also consider a particular and well-known example of the near-horizon deformations: Einstein-Gauss-Bonnet theory, where the regime of small Gauss-Bonnet coupling $\alpha$ is stable. We show that the spacing between modes for the perturbative (in $\alpha$) branch of modes is changing at an increasing rate when $n$ is growing. In addition, the new nonperturbative branch of modes appears even at the smallest coupling. As the lowest mode for this case was extensively studied, here we have complemented it by the analysis of the overtones for gravitational perturbations of Einstein-Gauss-Bonnet-AdS black holes in $D=5$ spacetime.

This work is organised as follows. Sec.~\ref{sec:param} develops the general parametrization of the asymptotically AdS black holes. Sec.~\ref{sec:masterequation} is devoted to the master wavelike equations for scalar and electromagnetic fields.
In Sec.~\ref{sec:methods} we briefly review the methods used for finding quasinormal modes. Sec.~\ref{sec:QNMs} describes features of the obtained quasinormal modes.
Finally, in Sec.~\ref{sec:conclusions} we summarize the obtained results and discuss relation of our work to other observations in this area.

\section{General parametrized spherically symmetric black hole in AdS space}\label{sec:param}

The powerful formalism for testing various properties of black holes in arbitrary metric theories of gravity was suggested for spherical and slowly rotating asymptotically flat black holes in \cite{Rezzolla:2014mua} and extended to the general axially symmetric case in \cite{Konoplya:2016jvv}. The essence of this approach is a general parametrization of the black-hole spacetime, similar in spirit with the post-Newtonian parametrized formalism, but working in the whole space outside the black hole including the near-horizon zone. This formalism was extensively applied for analytic approximations of various numerical black-hole metrics (see, for instance, \cite{Kokkotas:2017ymc,Kokkotas:2017zwt,Konoplya:2019fpy} and references therein) as well as for analysis of various radiation phenomena around them \cite{Younsi:2016azx,Konoplya:2023owh,Konoplya:2022tvv,Abdikamalov:2021zwv,Ni:2016uik,Konoplya:2021slg,Shashank:2021giy,Nampalliwar:2019iti,Konoplya:2021qll,Konoplya:2020jgt}. The advantage of this method is that frequently it allows one to approximate the black hole spacetime by only a few parameters for description of astrophysically relevant phenomena \cite{Konoplya:2020hyk}.

In this section the parametrization suggested initially for $D=4$ asymptotically de Sitter black holes in \cite{Konoplya:2022kld} is reviewed with the minimal specifications concerning the AdS asymptotic, while for more interesting $D=5$ case we generalize the approach of \cite{Konoplya:2020kqb} allowing now for a nonzero $\Lambda$-term.

\subsection{$D=4$}

Here we will use the general parametrization of spherically symmetric black holes in metric theories of gravity~\cite{Rezzolla:2014mua}, which was extended for the non asymptotically flat backgrounds in \cite{Konoplya:2022kld}.
The metric of a spherically symmetric black hole can be written in the following general form:
\begin{equation}
ds^2=-N^2(r)dt^2+\frac{B^2(r)}{N^2(r)}dr^2+r^2 (d\theta^2+\sin^2\theta d\phi^2),\label{metric}
\end{equation}
where $r_0$ is the event horizon, satisfying \mbox{$N(r_0)=0$.}

We introduce the dimensionless variable
\begin{equation}\label{xdef}
x \equiv 1-\frac{r_0}{r},
\end{equation}
so that $x=0$ corresponds to the event horizon, while \mbox{$x=1$} corresponds to spatial infinity. We define the function $N$ as
$$N^2=x A(x),$$
where $A(x)>0$ for \mbox{$0\leq x\leq1$}.

Following \cite{Konoplya:2022kld}, we represent the functions $A$ and $B$ as follows:
\begin{eqnarray}\label{Aexp}
A(x)&=&-\lambda(1-x)^{-2}-(\nu+\lambda)(1-x)^{-1}+\kappa
\\\nonumber&&-\epsilon (1-x)+(a_0-\epsilon)(1-x)^2+{\tilde A}(x)(1-x)^3\,,
\\
B(x)&=&1+b_0(1-x)+{\tilde B}(x)(1-x)^2\,,\label{Bexp}
\end{eqnarray}
where $\lambda$ and $\nu$ are the far-region asymptotic coefficients, corresponding to the cosmological constant,
$$\lambda=\frac{\Lambda r_0^2}{3},$$
and the effective dark-matter term~$\nu$~\cite{Mannheim:1988dj}. We also define
$$\kappa\equiv1-\lambda-\nu.$$

The parameter $\epsilon$ is related to the asymptotic mass $M$ in the following way:
$$\epsilon=\frac{2M}{r_0}-\kappa=\frac{2M}{r_0}-1+\lambda+\nu,$$
while the coefficients $a_0$ and $b_0$ correspond to the post-Newtonian parameters.

The functions ${\tilde A}$ and ${\tilde B}$ are introduced through the infinite continued fraction in order to describe the metric near the horizon (\ie for $x \simeq 0$),
\begin{equation}\label{ABdef}
{\tilde A}(x)=\dfrac{a_1}{1+\dfrac{a_2x}{1+\dfrac{a_3x}{1+\ldots}}}, \quad
{\tilde B}(x)=\dfrac{b_1}{1+\dfrac{b_2x}{1+\dfrac{b_3x}{1+\ldots}}},
\end{equation}
where $a_1, a_2,\ldots$ and $b_1, b_2,\ldots$ are dimensionless constants to be constrained from observations of phenomena which are localized near the event horizon. At the horizon only the first term in each of the continued fractions survives,
$
{\tilde A}(0)={a_1},~
{\tilde B}(0)={b_1},
$
which implies that near the horizon only the lower-order terms of the expansions are essential.

As we are aimed here only at the AdS configurations, further we will drop the dark-matter or post-Newtonian terms from the general parametrization, so that
$$\nu=a_0=b_0=0,$$
and the coefficient $\lambda$ is negative,
\begin{equation}\label{AdSradius}
\lambda=-\frac{r_0^2}{R^2},
\end{equation}
where $R$ is the AdS radius.

\subsection{$D=5$}

The higher-dimensional generalization for the spherically symmetric black hole parametrization has been proposed in \cite{Konoplya:2020kqb}. In a similar manner, we consider the line element
\begin{equation}
ds^2=-N^2(r)dt^2+\frac{B^2(r)}{N^2(r)}dr^2+r^2 d\Omega^2_{D-2},
\label{metricD}
\end{equation}
where $d\Omega^2_{D-2}$ is the metric of the unit $(D-2)$-sphere. The compact dimensionless variable ${\tilde x}$ is defined as follows:
\begin{equation}\label{xdefD}
{\tilde x} \equiv 1-\left(\frac{r_0}{r}\right)^{D-3},
\end{equation}

We notice that the consistent (anti)-de Sitter black hole representation in terms of the variable ${\tilde x}$, such that
$$N^2={\tilde x} A({\tilde x}),$$
is possible only for $D=5$. For the five-dimensional AdS black hole we need to choose
$$\lambda=0, \qquad \nu=-\frac{r_0^2}{R^2},$$
and the functions $A({\tilde x})$ and $B({\tilde x})$ are defined in the same way as for $D=4$, via Eq.~(\ref{ABdef}).

\subsection{Parametrization for the Gauss-Bonnet-AdS black hole}

The Lagrangian of the $D$-dimensional Einstein-Gauss-Bonnet theory \cite{Boulware:1985wk},
\begin{equation}\label{gbg3}
\mathcal{L}=-2\Lambda+R+\frac{\alpha}{2}(R_{\mu\nu\lambda\sigma}R^{\mu\nu\lambda\sigma}-4\,R_{\mu\nu}R^{\mu\nu}+R^2),
\end{equation}
where $\alpha$ is the Gauss-Bonnet coupling, leads to the equations allowing for a static spherically symmetric black-hole solution given by the line element (\ref{metricD}) with
\begin{eqnarray}
N^2(r)&=&1-\frac{\dfrac{4r^2}{D-2}\left(\dfrac{\mu}{r^{D-1}}+\dfrac{\Lambda}{D-1}\right)}{1+\sqrt{1+\dfrac{8\a}{D-2}\left(\dfrac{\mu}{r^{D-1}}+\dfrac{\Lambda}{D-1}\right)}},\nonumber\\\label{GBmetric}
B(r)&=&1, \qquad \widetilde{\alpha}\equiv\alpha\frac{(D-3)(D-4)}{2},
\end{eqnarray}
where $\mu$ is a constant, which defines the asymptotic mass and can be expressed in terms of the horizon radius, as follows \cite{Konoplya:2017zwo},
\begin{equation}\label{GBradius}
\mu=\frac{(D-2)r_0^{D-5}}{2} \left(r_0^2+ \widetilde{\alpha}- \frac{2 r_0^4 \Lambda}{(D-1)(D-2)} \right).
\end{equation}

Following \cite{Konoplya:2020kqb}, by comparing the asymptotic behavior of (\ref{GBmetric}) and the parametrized functions for $D=5$, we find the values of the asymptotic parameters:
\begin{eqnarray}
  \nonumber
  &&\nu=\frac{r_0^2 \Lambda}{3 + \sqrt{9 + 6 \alpha\Lambda}},
  \\\label{AGBpars}
  &&\epsilon=\frac{2\mu}{r_0^2\sqrt{9 + 6 \alpha\Lambda}}-1+\nu,
  \\\nonumber
  &&a_0=b_0=0.
\end{eqnarray}


Similarly, by comparing the near-horizon expansions for the function $N^2(r)$, one can calculate the near-horizon coefficients,
\begin{eqnarray}
    a_1&=&2 \nu -\frac{r_0^4 \Lambda -6 r_0^2 \epsilon -12 \alpha  \epsilon +6 \alpha }{3 r_0^2+6\alpha},\\
    a_2&=&\frac{r_0^8 \Lambda  (3-2 \alpha  \Lambda )}{18 a_1 \left(r_0^2+2 \alpha \right)^3}+\frac{r_0^6 (2 \alpha  \Lambda +3 \epsilon -\nu )}{a_1 \left(r_0^2+2 \alpha \right)^3}
    \\\nonumber&&
    +\frac{r_0^4 \alpha  (2 \alpha \Lambda +18 \epsilon -6 \nu -3)}{a_1 \left(r_0^2+2 \alpha \right)^3}-3
    \\\nonumber&&
   +\frac{4 r_0^2 \alpha ^2 (9 \epsilon -3 \nu -1)+8 \alpha ^3 (3 \epsilon -\nu )}{a_1 \left(r_0^2+2 \alpha \right)^3},\ldots
\end{eqnarray}
and $b_1=0$. Notice that when finding quasinormal modes of the Einstein-Gauss-Bonnet-AdS black holes with the Bernstein polynomial method, we use the full metric and the truncated parametrization is necessary here to show that the Gauss-Bonnet correction belong to the considered class of near-horizon deformations. If one uses Frobenius series approach for finding quasinormal modes (\eg Horowitz-Hubeny method \cite{Horowitz:1999jd}), then the parametrization must be extended to a sufficiently high order for which the quasinormal frequencies approach the accurate ones for the full metric, as was done, for example, in \cite{Konoplya:2022iyn}.

\section{Wave equation}\label{sec:masterequation}

Once we are illustrating general idea of the near-horizon deformation of the black-hole geometry rather than fix any particular theory of gravity, we may have two approaches to perturbation equations. The first way is to consider test fields in a black hole background and deform the metric near the horizon, looking at the consequent changes in the spectrum. The second approach is to consider master wave equations for gravitational perturbations of the Schwarzschild-AdS solution and implement small near horizon deformations directly therein. We will use both approaches here.

The general covariant equations for the scalar ($\Phi$) and electromagnetic ($A_\mu$) fields can be written in the following way:
\begin{equation}\label{perturb}
\begin{array}{lcl}
\dfrac{1}{\sqrt{-g}}\partial_\mu \left(\sqrt{-g}g^{\mu \nu}\partial_\nu\Phi\right)&=&0,
\\
\dfrac{1}{\sqrt{-g}}\partial_{\mu} \left(F_{\rho\sigma}g^{\rho \nu}g^{\sigma \mu}\sqrt{-g}\right)&=&0\,,
\end{array}
\end{equation}
where $F_{\mu\nu}=\partial_\mu A_\nu-\partial_\nu A_\mu$ is the electromagnetic tensor.

After separation of the variables the above equations (\ref{perturb}) take the Schrödinger wavelike form
\begin{equation}\label{wavelike}%
\left(\frac{d^2}{dr_*^2} + \omega^2 - V(r_*)\right)\Psi(r_*) = 0,
\end{equation}
with the effective potential,
\begin{equation}\label{potantial}
V(r) = N^2(r)\frac{\ell(\ell+1)}{r^2} + \frac{1-s}{r}\frac{d}{dr_*}\frac{N^2(r)}{B(r)},
\end{equation}
for the scalar ($s=0$) and electromagnetic ($s=1$) fields, and the tortoise coordinate is defined as
\begin{equation}\label{tortoise}
dr_*=\frac{B(r)dr}{N^2(r)}.
\end{equation}

Equation (\ref{wavelike}) has two regular singular points, $r=r_0$ and $r=\infty$, and the appropriate boundary conditions are the ingoing wave at the horizon and vanishing at infinity,
\begin{eqnarray}\label{ingoing}
&&\Psi \propto (r-r_0)^{-\imo\omega/2\kappa_0}, \quad r\to r_0, \\
\label{Dirichlet}
&&\Psi \propto r^{-s-(1-s)D/2}, \qquad r\to\infty,
\end{eqnarray}
where $\kappa_0\equiv N(r_0)N'(r_0)/B(r_0)$ is the surface gravity at the horizon.

\section{The methods for studying quasinormal spectrum}\label{sec:methods}

Here we will briefly summarize two methods used for finding quasinormal modes: the Bernstein polynomial method and the shooting method. Both approaches are in excellent agreement, convergent and allow one to control the accuracy of the obtained results.

\subsection{Bernstein spectral method}\label{sec:Bernstein}

Following \cite{Fortuna:2020obg}, we introduce the function $\phi(x)$, which is regular for $0\leq x\leq 1$ when $\omega$ is a quasinormal mode,
\begin{equation}\label{regularized}
\Phi(r)=\left(1-\frac{r_0}{r}\right)^{-\imo\omega/2\kappa_0}r^{-s-\frac{(1-s)D}{2}}\phi\left(1-\frac{r_0}{r}\right),
\end{equation}
and represent $\phi(x)$ as a sum
\begin{equation}\label{Bernsteinsum}
\phi(x)(1-x)^2=\sum_{k=0}^{N-2}C_kB_k^N(x),
\end{equation}
where
$$B_k^N(x)\equiv\frac{N!}{k!(N-k)!}x^k(1-x)^{N-k}$$
are the Bernstein polynomials.

The representation (\ref{Bernsteinsum}) is chosen in order to satisfy the Dirichlet condition at the AdS boundary (\ref{Dirichlet}), which provides the best accuracy of the Bernstein approximation \cite{Konoplya:2022xid}. We used the Mathematica\textregistered{} code publicly shared in \cite{Konoplya:2022zav}.

\subsection{The shooting method}

In order to check the frequencies obtained via Bernstein polynomial approximation we perform numerical integration of the wavelike differential equation. We introduce the new function, which is finite at the AdS bound ($x=1$),
\begin{equation}
f(x)=(1-x)^{s+(D-1)(1-s)}\phi(x),
\end{equation}
then, from the wavelike equation (\ref{wavelike}) we find a Taylor expansion for the function $f(x)$, assuming that it is regular at the horizon ($x=0$). We use the obtained series expansion as the initial approximation for the value of the function and its derivative at some point near the horizon $x=x_0\ll1$, where we set up the initial condition for the second-order differential equation for $f(x)$. For each value of $\omega$ we solve the equation numerically using the Runge-Kutta method in the interval $x_0\leq x\leq1$ and find numerically the value $f(1)$. We take $x_0=0.05$ and make sure that the approximation given by the series expansion in this point is sufficient within the chosen tolerance of the numerical solution.

The Dirichlet boundary condition (\ref{Dirichlet}) is equivalent to
\begin{equation}\label{regDirichlet}
f(1)=0.
\end{equation}
In order to obtain the quasinormal frequency we shoot for the value of $\omega$ until we satisfy the condition (\ref{regDirichlet}). This method allows us to calculate the dominant quasinormal frequencies.

\section{Quasinormal modes}\label{sec:QNMs}

\begin{table*}
\begin{tabular}{|c|c|c|c|c|}
  \hline
  n & $a_{1}=0$ & $a_{1}=10^{-4}$ & $a_{1}=10^{-3}$ & $a_{1}=10^{-2}$\\
\hline
  0 & ~2.798223 - ~2.671206 i
    & ~2.817861 - ~2.669980 i
    & ~2.976585 - ~2.638038 i
    & ~3.689333 - ~2.241428 i \\
    & ~2.798223 - ~2.671206 i
    & ~2.817861 - ~2.669980 i
    & ~2.976585 - ~2.638038 i
    & ~3.689333 - ~2.241428 i \\
\hline
  1 & ~4.758489 - ~5.037569 i
    & ~4.867591 - ~5.041058 i
    & ~5.429435 - ~4.851398 i
    & - 2.970848363917849 i \\
    & ~4.758489 - ~5.037569 i
    & ~4.867591 - ~5.041058 i
    & ~5.429435 - ~4.851398 i
    & - 2.970848363917732 i\\
\hline
  2 & ~6.719268 - ~7.394493 i
    & ~7.004002 - ~7.374227 i
    &  - 5.514253127354089 i
    & ~6.906420 - ~4.068041 i \\
    & ~6.719268 - ~7.394493 i
    & ~7.004002 - ~7.374227 i
    &  - 5.514253127354271 i
    & ~6.906420 - ~4.068041 i \\
\hline
  3 & ~8.682227 - ~9.748517 i
    & ~9.181600 - ~9.642673 i
    & ~7.893864 - ~6.974041 i
    & 10.090392 - ~5.882584 i \\
    & ~8.682227 - ~9.748517 i
    & ~9.181600 - ~9.642673 i
    & ~7.893864 - ~6.974041 i
    & 10.090392 - ~5.882584 i \\
\hline
  4 & 10.646670 - 12.101248 i
    & - 11.092746111436081 i
    & 10.344339 - ~9.092176 i
    & 13.265095 - ~7.687109 i \\
    & 10.646670 - 12.101248 i
    & - 11.092746111436082 i
    & 10.344339 - ~9.092176 i
    & 13.265095 - ~7.687109 i \\
\hline
  5 & 12.612107 - 14.453276 i
    & 11.353990 - 11.867664 i
    & 12.793191 - 11.209959 i
    & 16.435264 - ~9.486718 i \\
    & 12.612107 - 14.453277 i
    & 11.353990 - 11.867664 i
    & 12.793223 - 11.209959 i
    & 16.435277 - ~9.486721 i \\
\hline
  6 & 14.578238 - 16.804876 i
    & 13.511339 - 14.076038 i
    & 15.241362 - 13.325540 i
    & 19.602705 - 11.283490 i \\
    & 14.578269 - 16.804821 i
    & 13.511339 - 14.076038 i
    & 15.242962 - 13.328813 i
    & 19.603995 - 11.283287 i \\
\hline
  7 & 16.544871 - 19.156190 i
    & 15.657220 - 16.280837 i
    & - 15.4152947700877 i
    & - 11.3124664110 i \\
    & 16.521359 - 19.179846 i
    & 15.657235 - 16.280869 i
    & - 15.4152947686748 i
    & - 11.3124664102 i \\
\hline
  8 & 18.511881 - 21.507304 i
    & 17.796306 - 18.486619 i
    & 17.688451 - 15.439313 i
    & 22.768315 - 13.078455 i \\
\hline
  9 & 20.479182 - 23.858271 i
    & 19.931921 - 20.694265 i
    & 20.13456~ - 17.55187~ i
    & 25.932606 - 14.872186 i \\
\hline
\end{tabular}
\caption{QNMs for $s=\ell=0$, $\epsilon=a_{0}=b_{0}=0$, $r_{0}=R=1$, $a_{2}=-1000$, $a_{3}=1001$, $a_{i}=0$ for $i>3$. First line: Bernstein spectral method, second line: shooting method.}\label{tabl:r01}
\end{table*}

\begin{figure*}
\centerline{\resizebox{\linewidth}{!}{\includegraphics*{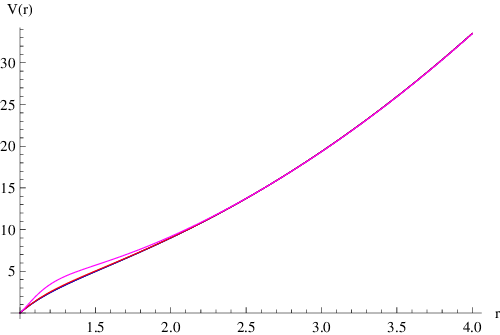}\includegraphics*{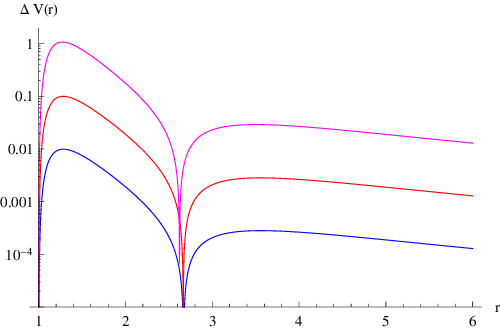}}}
\caption{Left panel: the effective potentials for Schwarzschild-AdS and two types of deformed metrics. Right panel: the difference between the effective potentials of the nondeformed (SAdS) and deformed metrics; $\ell=s=\epsilon=a_{0}=0$, $R=1$, $r_{0}=1$, $a_{1}=10^{-4}$ (bottom,blue), $a_{1}=10^{-3}$ (middle, red), $a_{1}=10^{-2}$ (top, magenta) with $a_{2}=-1000$, $a_{3}=1001$, $a_{i}=0$ for $i>3$.}\label{fig:SAdSdeformation}
\end{figure*}

We will start from the four-dimensional example of the Schwarzschild-AdS black hole and compare it with the metric which is deformed near the event horizon but the deformation slowly decay at larger $r$, so that at about $r\approx 3 r_{0}$ the difference between the deformed and original spacetime is negligible. This way we create a deformation which is not highly localized and would not require high energies to support it. This kind of smooth deformation in some spread region rather than the highly localized deformation is also required by realistic physical configurations: the fields, even when they decay quickly, exist everywhere in space and must be smooth.

The quasinormal modes of asymptotically AdS black holes are different depending on the ratio between the radius of the event horizon $r_{0}$ and the AdS radius $R$. We can conditionally distinguish the three regimes.
\begin{enumerate}
\item When $r_{0} \ll R$ (small black holes), the spectrum usually resembles the one of the pure AdS spacetime, approaching normal modes of the AdS space in the limit $r_{0} \to 0$ \cite{Konoplya:2002zu}, provided such small black holes are dynamically stable in the theory under consideration.
\item When $r_{0} \gg R$ (large black holes) the highly damped modes quickly achieve the regime of equidistant spacing. The regime of large black holes is the most important for the AdS/CFT correspondence as one can think the large black hole is similar to the black brane in this regime.
\item When $r_{0} \sim R$ (intermediate black holes) we have a transition between the pure AdS spacetime and large black holes regime.
\end{enumerate}

From table~\ref{tabl:r01} we can see the quasinormal modes of the test scalar field for $\ell=0$ perturbations of the intermediate black holes $r_{0}=R=1$. The data obtained here by the Bernstein polynomial and shooting methods are in excellent agreement with that obtained by the Horowitz-Hubeny method \cite{Horowitz:1999jd,Konoplya:2002zu}. There one can see that all the frequencies have nonzero real and imaginary parts and the modes approaching the equidistant spectrum as the overtone number is increased.

We also present quasinormal frequencies for small deformations of the metric function near the event horizon shown in Fig.~\ref{fig:SAdSdeformation}. The bigger value of $a_1$ corresponds to the larger deformation. We can see that the spacing between the nearby frequencies is changing at much larger rate when the deformation is turned on. Moreover, a new branch of modes on the imaginary axis appears in the spectrum. These modes completely change the picture of poles of the Green functions, as even at relatively small deformations they may appear already in the place of the first few low-lying modes. For instance, already for $a_1=10^{-3}$, the new branch of modes starts at the second overtone, while for $a_1=10^{-2}$ the purely imaginary mode is the first overtone.

In the limit of vanishing black hole radius $r_0$ (or mass $\mu$) the contribution of the Gauss-Bonnet correction does not vanish, which happens for an asymptotically flat case. Small black holes with $r_0^2\ll |\a|$ are unstable against gravitational perturbations as was shown, for example, in \cite{Konoplya:2017ymp}. Consequently, the quasinormal spectrum does not approach the purely AdS one as there is a gap in the range of $r_0$ in which the instability breaks down the black hole configuration.

At the same time, when there is no black hole ($\mu=0$), the spacetime is purely AdS, with the radius $R$ depending on $\a$,
$$\frac{1}{R^2}=-\dfrac{4\Lambda}{(D-1)(D-2)}\left(1+\sqrt{1+\dfrac{8\Lambda\a}{(D-1)(D-2)}}\right)^{-1}$$
and the spectrum of the normal modes for the gravitational perturbations in the purely AdS spacetime does not contain unstable modes \cite{Natario:2004jd},
\begin{equation}
\omega R = D + \ell - j + 2n., \qquad n\in\mathbb{N}.
\end{equation}
Here $j=1$ for tensor-type perturbations, $j=2$ vector-type perturbations, and $j=3$ for scalar-type perturbations.

In \cite{Arean:2023ejh} highly localized deformations were considered for the asymptotically flat spacetimes, which, in our opinion, somewhat misleadingly, were called ``high frequency perturbations'' and the correspondingly changed quasinormal spectrum was called ``pseudospectrum''. Those kind of deformations, although interesting mathematically, evidently cannot represent a situation we are considering here, when the deformations are supposed to be induced by adding extra matter fields, or curvature terms, or other physically or holographically motivated corrections.

\begin{table*}
\begin{tabular}{|c|c|c|c|c|}
  \hline
  \multicolumn{5}{|c|}{Scalar-type gravitational perturbations}\\
  \hline
  n & $ \alpha=-0.05$ & $\alpha=0$  & $\alpha=0.05$ & $\alpha=0.08$ \\
  \hline
  0 & ~2.47376 - ~0.54029 i
    & ~2.20478 - ~0.58137 i
    & ~1.95912 - ~0.57286 i
    & ~1.83306 - ~0.55220 i \\
  1 & ~5.24255 - ~2.30332 i
    & ~4.80507 - ~2.58486 i
    & ~4.61348 - ~2.69809 i
    & ~4.65242 - ~2.77405 i \\
  2 & ~8.19528 - ~3.96873 i
    & ~7.54268 - ~4.60733 i
    & ~7.52654 - ~4.88161 i
    & ~7.78514 - ~5.06530 i \\
  3 & 11.13940 - ~5.53812 i
    & 10.32318 - ~6.62130 i
    & 10.57627 - ~7.09311 i
    &  - 7.19970847625683 i \\
  4 & 14.05506 - ~7.06896 i
    & 13.12149 - ~8.63064 i
    & 13.67306 - ~9.32189 i
    & 10.99732 - ~7.41080 i \\
  5 &  - ~7.763361332275~ i
    & 15.92897 - 10.63734 i
    &  - 11.467802046439~ i
    & 14.23492 - ~9.81674 i \\
  6 & 16.94930 - ~8.58429 i
    & 18.74183 - 12.64241 i
    & 16.78659 - 11.57317 i
    & 17.49606 - 12.26552 i \\
  7 & 19.82890 - 10.09266 i
    & 21.5580~ - 14.6460~ i
    & 19.9111~ - 13.8466~ i
    & 20.7780~ - 14.7410~ i \\
  \hline
  \hline
  \multicolumn{5}{|c|}{Vector-type gravitational perturbations}\\
  \hline
  n & $ \alpha=-0.05$ & $\alpha=0$  & $\alpha=0.05$ & $\alpha=0.08$ \\
  \hline
  0 & ~5.35720 - ~2.41565 i
    & - 1.45861795797447 i
    & - 0.93899127219167 i
    & - 0.74452594828788 i \\
  1 & ~1.08405 - ~3.43947 i
    & ~4.99634 - ~2.36133 i
    & ~4.98538 - ~2.30107 i
    & ~5.04502 - ~2.31381 i \\
  2 & ~8.37330 - ~4.09124 i
    & ~7.70523 - ~4.47971 i
    & ~7.76948 - ~4.55582 i
    & ~7.96310 - ~4.71901 i \\
  3 & 11.29271 - ~5.63807 i
    & 10.46719 - ~6.52871 i
    & 10.71453 - ~6.82252 i
    & 11.07063 - ~7.16093 i \\
  4 & 14.17626 - ~7.15417 i
    & 13.25125 - ~8.55642 i
    & 13.74639 - ~9.10558 i
    & 14.26052 - ~9.62974 i \\
  5 & 17.04381 - ~8.65916 i
    & 16.04743 - 10.57460 i
    & 16.82385 - 11.39979 i
    & -10.34203287425 i \\
  6 & 19.90281 - 10.15951 i
    & 18.85112 - 12.58761 i
    & 19.92740 - 13.70460  i
    & 17.49616 - 12.12033 i \\
  7 & 22.75649 - 11.65775 i
    & 21.660~~ - 14.597~~ i
    & -15.08781874129 i
    & 20.763~~ - 14.626~~ i \\
  \hline
  \hline
  \multicolumn{5}{|c|}{Tensor-type gravitational perturbations}\\
  \hline
  n & $ \alpha=-0.05$ & $\alpha=0$  & $\alpha=0.05$ & $\alpha=0.08$ \\
  \hline
  0 & ~6.11794 - ~2.21602 i
    & ~5.79660 - ~2.16599 i
    & ~5.75896 - ~1.90920 i
    & ~5.83169 - ~1.79812 i \\
  1 & ~8.89360 - ~4.02117 i
    & ~8.32497 - ~4.28183 i
    & ~8.32192 - ~4.05824 i
    & ~8.45714 - ~4.07714 i \\
  2 &  - 5.65141597938718 i
    & 10.98254 - ~6.35001 i
    & 11.09063 - ~6.28957 i
    & 11.31138 - ~6.52339 i \\
  3 & 11.68540 - ~5.66341 i
    & 13.69816 - ~8.39604 i
    & 13.97285 - ~8.58568 i
    & 14.31711 - ~9.07436 i \\
  4 & 14.47585 - ~7.23263 i
    & 16.44540 - 10.42946 i
    & 16.93406 - 10.92479 i
    & 17.44440 - 11.67248 i \\
  5 & 17.27152 - ~8.76309 i
    & 19.21200 - 12.45496 i
    & 19.95573 - 13.28841 i
    & 20.66074 - 14.27511 i \\
  6 & 20.07573 - 10.27277 i
    & 21.99144 - 14.47515 i
    & 23.02387 - 15.66275 i
    & 23.93242 - 16.86341 i \\
  7 & 22.88813 - 11.77188 i
    & 24.77988 - 16.49158 i
    & 26.12660 - 18.03909 i
    & 27.23566 - 19.43522 i \\
  \hline
\end{tabular}
\caption{QNMs for the gravitational perturbations ($\ell=2$) of the five-dimensional Gauss-Bonnet black holes ($r_0=R=1$).}\label{tabl:Gauss-Bonnet-r1}
\end{table*}

\begin{table*}
\begin{tabular}{|c|c|c|c|c|}
  \hline
  \multicolumn{5}{|c|}{Scalar-type gravitational perturbations}\\
  \hline
  n & $ \alpha=-0.05$ & $\alpha=0$  & $\alpha=0.05$ & $\alpha=0.08$ \\
  \hline
  0 & ~~1.64197 - ~0.09574 i
    &  ~1.63975 - ~0.08305 i
    &  ~1.63770 - ~0.07081 i
    &  ~1.63655 - ~0.06369 i \\
  1 & ~33.07867 - 27.47543 i
    &  31.38586 - 27.44097 i
    &  30.84532 - 27.04797 i
    &  31.01640 - 26.92059 i \\
  2 & ~55.76607 - 46.38915 i
    &  51.97484 - 47.61518 i
    &  52.20465 - 47.99100 i
    &  53.63586 - 48.94729 i \\
  3 & ~78.04014 - 64.04380 i
    &  72.25286 - 67.67731 i
    &  74.01045 - 69.57613 i
    &  76.95270 - 71.73247 i \\
  4 & - 74.39673029 i
    &  92.44166 - 87.70772 i
    &  96.2318~ - 91.4897~ i
    &  - 92.869 i \\
  5 & ~99.82944 - 81.13181 i
    &  ~112.594 - 107.726 i
    & $^\dag$~~ -433.485714957 i
    &  100.443 - 94.7882 i \\
  \hline
  \hline
  \multicolumn{5}{|c|}{Vector-type gravitational perturbations}\\
  \hline
  n & $ \alpha=-0.05$ & $\alpha=0$  & $\alpha=0.05$ & $\alpha=0.08$ \\
  \hline
  0 &  -0.1446909840048998 i
    & - 0.1252395152661459 i
    & - 0.1065241589233073 i
    & - 0.0956857407606597 i \\
  1 & ~33.12108 - ~27.47324 i
    & ~31.42895 - ~27.42769 i
    & ~30.89095 - ~27.01957 i
    & ~31.06271 - ~26.88247 i \\
  2 & ~55.79525 - ~46.39269 i
    & ~52.00549 - ~47.60653 i
    & ~52.23675 - ~47.96177 i
    & ~53.66123 - ~48.90858 i \\
  3 & ~78.06157 - ~64.04865 i
    & ~72.27743 - ~67.67056 i
    & ~74.03081 - ~69.54649 i
    & ~76.96189 - ~71.70095 i \\
  4 & - 74.334587202459727 i
    & ~92.46250 - ~87.70206 i
    & ~96.24229 - ~91.46318 i
    & - 93.24616226105020597 i \\
  5 & ~99.84551 - ~81.13702 i
    & 112.61191 - 107.72042 i
    & 118.64084 - 113.50264 i
    & 100.44518 - ~94.76488 i \\
  6 & 121.29630 - ~98.01024 i
    & 132.74359 - 127.73230 i
    & - 134.12834960251351 i
    & ~123.9424 - 118.09517 i \\
  7 & 142.53479 - 114.79737 i
    & 152.86531 - 147.74055 i
    & 141.07865 - 135.64126 i
    & 147.49294 - 141.65175 i \\
  \hline
  \hline
  \multicolumn{5}{|c|}{Tensor-type gravitational perturbations}\\
  \hline
  n & $ \alpha=-0.05$ & $\alpha=0$  & $\alpha=0.05$ & $\alpha=0.08$ \\
  \hline
  0 & ~33.24895 - ~27.46565 i
    & ~31.55838 - ~27.38851 i
    & ~31.02709 - ~26.93635 i
    & ~31.20066 - ~26.77078 i \\
  1 & ~55.88289 - ~46.40219 i
    & ~52.09753 - ~47.58078 i
    & ~52.33247 - ~47.87540 i
    & ~53.73739 - ~48.79396 i \\
  2 & ~78.12577 - ~64.06252 i
    & ~72.35121 - ~67.65040 i
    & ~74.09182 - ~69.45850 i
    & ~76.98978 - ~71.60676 i \\
  3 & - 74.162239894658740 i
    & ~92.52504 - ~87.68514 i
    & ~96.27417 - ~91.38387 i
    & - 94.373003543864476 i \\
  4 & ~99.89361 - ~81.15222 i
    & 112.66671 - 107.70566 i
    & 118.65471 - 113.43699 i
    & 100.45168 - ~94.69473 i \\
  5 & 121.33310 - ~98.02553 i
    & 132.79268 - 127.71912 i
    & - 135.419747021130999 i
    & 123.93986 - 118.04132 i \\
  6 & 142.56346 - 114.81214 i
    & 152.90997 - 147.72858 i
    & 141.08308 - 135.58775 i
    & 147.48560 - 141.60900 i \\
  7 & 163.65593 - 131.56094 i
    & 173.02204 - 167.73557 i
    & 163.52222 - 157.85064 i
    & 171.11181 - 165.32335 i \\
  \hline
\end{tabular}
\caption{QNMs for the gravitational perturbations ($\ell=2$) of the five-dimensional Gauss-Bonnet black holes ($r_0=10R=10$). $^\dag$ The pure damped (nonoscillatory) mode for small $\alpha$ in scalar-type channel corresponds to a large value of the overtone: we have shown it for $\alpha=0.05$ at $n=5$ position in the place of the corresponding mode with nonzero real and imaginary parts.}\label{tabl:Gauss-Bonnet-r10}
\end{table*}

Indeed, a particular example we consider here in tables~\ref{tabl:Gauss-Bonnet-r1} and~\ref{tabl:Gauss-Bonnet-r10} is the five-dimensional Einstein-Gauss-Bonnet theory, which is the most interesting case, because it is dual to the four-dimensional field theory. Here the coupling $\alpha$ is considered to be very small, so that the fundamental mode deviates insignificantly from its Schwarzschild values. However, we see that the new branch of modes appears in that case. This effect was extensively studied in~\cite{Konoplya:2017zwo,Konoplya:2017ymp,Gonzalez:2017gwa}, but essentially for the slowest damping modes. In~\cite{Konoplya:2017zwo} it was shown that once $r_0/R \gtrapprox \ell$ the least damped sound mode in the vector channel of gravitational perturbations obeys a simple fit,
\begin{eqnarray}\label{QNMhydro}
\omega = -\frac{(\ell-1)(\ell+D-2)}{(D-1)r_0}\left(1-\frac{D+1}{D-3}\cdot\frac{\a}{R^2}\right)\imo,
\end{eqnarray}
where $\a = \alpha (D-3) (D-4)/2$. Here we will see that in addition to this purely imaginary mode, there is a full branch of nonoscillatory modes, which, however, should not be confused with the sound mode.

\begin{figure*}
\centerline{\resizebox{\linewidth}{!}{\includegraphics*{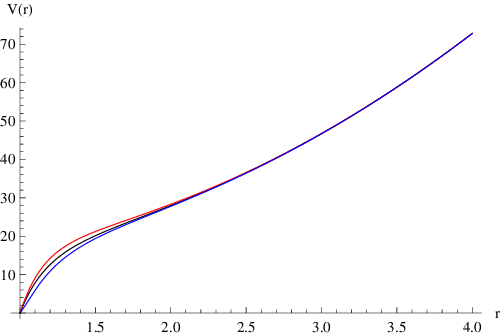}\includegraphics*{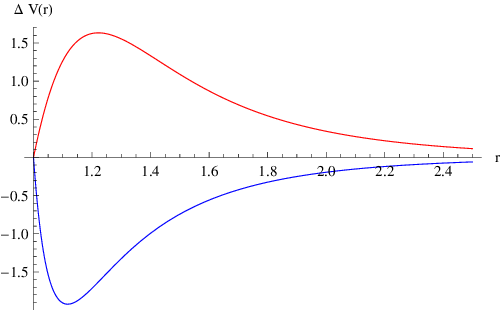}}}
\caption{Left panel: the effective potentials for the tensor-type perturbations ($\ell=2$) of the 5D Gauss-Bonnet-AdS black hole ($r_{0}=R=1$) for $\alpha=0.05$ (red, upper) and $\alpha=-0.05$ (blue, lower) and 5D Tangherlini-AdS black hole ($\alpha=0$) for comparison (black). Right panel: the difference between the effective potentials of the Gauss-Bonnet-AdS (deformed metric) and Tangherlini-AdS (nondeformed metric).}\label{fig:GBpotentials}
\end{figure*}

Therefore, here by the revision of the spectrum for the Einstein-Gauss-Bonnet black holes we have in mind two purposes. First, we present the quasinormal modes for this case as an illustration of the fact that the small deformations we consider in this paper have a physically motivated origin and small coupling $\alpha$ plays the role of such a near-horizon deformation parameter (see Fig.~\ref{fig:GBpotentials}). The second aim is to complete the previous study of the Einstein-Gauss-Bonnet-AdS quasinormal spectrum via the analysis of its overtones.

The gravitational perturbations can be treated independently for the three channels which transforms as tensors, vectors and scalars relatively the three-dimensional sphere. The appropriate master wave equations can be found in \cite{Dotti:2005sq,Gleiser:2005ra} and also summarized in the appendix to this article for convenience.
From table~\ref{tabl:Gauss-Bonnet-r1} one can see the new purely imaginary mode in the spectrum for all three channels of gravitational perturbations of the intermediate-size black holes ($r_{0} =R$). However, for the tensor channel it appears for negative $\alpha$. These new nonoscillatory modes should not be confused with the fundamental purely imaginary mode of the vector type of gravitational perturbations, responsible for the hydrodynamic pole in the dual field theory. The position of this fundamental mode agrees with the results of \cite{Konoplya:2017zwo,Konoplya:2017ymp}. These new purely imaginary modes are nonperturbative on the coupling $\alpha$ and were discussed first in the context of the holographic models of thermalization \cite{Grozdanov:2016vgg}.

The quasinormal modes of large black holes ($r_{0}=10 R$) in Einstein-Gauss-Bonnet-AdS theory are presented in table \ref{tabl:Gauss-Bonnet-r10}. There one can see that the new purely imaginary modes appear at slightly higher overtone numbers $n$.

For all the considered cases, we observe a specific consequence of the small near-horizon deformations: the presence of the new branch of modes changes the structure of the overtones seemingly. Therefore, once even a small correction is considered on the AdS gravity side of the correspondence, the spectrum must carefully revised, because it may considerably change the structure of the poles in the dual field theory

In the course of calculations, we have also observed that the Bernstein polynomial method is a powerful and automatic tool for finding quasinormal modes of the asymptotically AdS black holes, especially of the purely imaginary modes which may be important for the dual description of hydrodynamics.

\section{Conclusions}\label{sec:conclusions}

Quasinormal modes of asymptotically AdS black holes are important, because they may be interpreted as poles of the retarded Green functions in the dual Conformal Field Theory and describe strongly coupled systems, such as quark-gluon plasma, which cannot be treated by the usual perturbative methods of the Quantum Field Theory.

The phenomenon we studied here is the high sensitivity of overtones (in comparison with the fundamental mode) to the least deformations of the geometry in some region near the event horizon. Such small deformations may come, for instance, from higher curvature corrections to the Einstein action, which is used to understand the intermediate coupling regime in the dual field theory.

Here we have obtained the following results:
\begin{itemize}
\item We have shown that ad hoc deformation of the Schwarzschild-AdS black hole in some region near the event horizon leads to almost the values of the fundamental modes.
This feature of the quasinormal spectrum is similar to the asymptotically flat case, where, for small deformations of the effective potential in the whole region outside the event horizon, the fundamental mode is destabilized at the same order as the size of the deformation \cite{Cheung:2021bol,Berti:2022xfj}. When small deformations of the asymptotically flat black holes are restricted by the near-horizon zone the fundamental mode is practically indistinguishable from the Schwarzschild/Kerr limit \cite{Konoplya:2022pbc,Konoplya:2023hqb}.
\item The near-horizon deformations drastically change the picture of overtones, leading even to the appearance of a new branch of nonperturbative (in the Gauss-Bonnet coupling $\alpha$) modes with vanishing real parts. This change of the overtones' structure may occur not only at high overtones, but already at $n=1$ for relatively small near-horizon deformations. The spacing of the perturbative branch changes at an increasing rate when the overtone number $n$ is increased.
\item In order to prepare deformation solely in the near horizon zone, we extended the general parametrization of the asymptotically flat black holes in four and five spacetime dimensions to the asymptotically AdS case. This parametrization can be further applied to finding analytical approximations for the numerical asymptotically-AdS black holes solutions.
\item As a well-known example of the above small deformations we have considered gravitational perturbations of the Einstein-Gauss-Bonnet-AdS black holes and studied the overtones behavior for this case in detail.
\end{itemize}

Our work on the black hole parametrization could be extended to the case of rotating asymptotically AdS black holes using similar approach which was developed for the asymptotically flat case \cite{Konoplya:2016jvv}. However, the parametrization in higher than five dimensions would require a different ansatz to include the lower dimensional black holes as well.

\acknowledgments
A.~Z. was supported by Conselho Nacional de Desenvolvimento Científico e Tecnológico (CNPq).

\appendix
\section*{Appendix}

Here we summarize the basics of perturbations of the gravitational sector in the Einstein-Gauss-Bonnet theory allowing for a cosmological constant.
After decoupling of the angular variables the gravitational perturbation equations can be reduced to the second-order master differential equations \cite{Takahashi:2010ye}
\begin{equation}
\left(\frac{\partial^2}{\partial t^2}-\frac{\partial^2}{\partial r_*^2}+V_i(r_*)\right)\Psi(t,r_*)=0,
\end{equation}
where $r_*$ is the tortoise coordinate,
\begin{equation}
dr_*\equiv \frac{dr}{N^2(r)}=\frac{dr}{1-r^2\psi(r)},
\end{equation}
and $i$ stands for $t$ (\emph{tensor}), $v$ (\emph{vector}), and $s$ (\emph{scalar}) \
types of gravitational perturbations.
The explicit forms of the effective potentials $V_s(r)$, $V_v(r)$, and $V_t(r)$ \cite{Cuyubamba:2016cug} are given by
\begin{eqnarray}\label{potentials}
V_t(r)&=&\frac{\ell(\ell+n-1)N^2(r)T''(r)}{(n-2)rT'(r)}+\frac{1}{R(r)}\frac{d^2R(r)}{dr_*^2},\\\nonumber
V_v(r)&=&\frac{(\ell-1)(\ell+n)N^2(r)T'(r)}{(n-1)rT(r)}+R(r)\frac{d^2}{dr_*^2}\Biggr(\frac{1}{R(r)}\Biggr),\\\nonumber
V_s(r)&=&\frac{2\ell(\ell+n-1)N^2(r)P'(r)}{nrP(r)}+\frac{P(r)}{r}\frac{d^2}{dr_*^2}\left(\frac{r}{P(r)}\right),
\end{eqnarray}
where $n\equiv D-2$, $\ell=2,3,4,\ldots$ is the multipole number and functions $T(r)$, $R(r)$, and $P(r)$ can be written as follows
\begin{eqnarray}\nonumber
T(r)&=& =\frac{nr^{n-1}}{2}\Biggr(1+2\a\psi(r)\Biggr),
\\
R(r)&=&r\sqrt{T'(r)},
\\\nonumber
P(r)&=&\frac{2(\ell-1)(\ell+n)-nr^3\psi'(r)}{\sqrt{T'(r)}}T(r).
\end{eqnarray}

\bibliographystyle{unsrt}
\bibliography{bibliography}

\begin{thebibliography}{10}

\bibitem{Maldacena:1997re}
Juan~Martin Maldacena.
\newblock {The Large N limit of superconformal field theories and
  supergravity}.
\newblock {\em Adv. Theor. Math. Phys.}, 2:231--252, 1998.

\bibitem{Gubser:1998bc}
S.~S. Gubser, Igor~R. Klebanov, and Alexander~M. Polyakov.
\newblock {Gauge theory correlators from noncritical string theory}.
\newblock {\em Phys. Lett. B}, 428:105--114, 1998.

\bibitem{Witten:1998qj}
Edward Witten.
\newblock {Anti-de Sitter space and holography}.
\newblock {\em Adv. Theor. Math. Phys.}, 2:253--291, 1998.

\bibitem{Birmingham:2001pj}
Danny Birmingham, Ivo Sachs, and Sergey~N. Solodukhin.
\newblock {Conformal field theory interpretation of black hole quasinormal
  modes}.
\newblock {\em Phys. Rev. Lett.}, 88:151301, 2002.

\bibitem{Son:2002sd}
Dam~T. Son and Andrei~O. Starinets.
\newblock {Minkowski space correlators in AdS / CFT correspondence: Recipe and
  applications}.
\newblock {\em JHEP}, 09:042, 2002.

\bibitem{Kovtun:2005ev}
Pavel~K. Kovtun and Andrei~O. Starinets.
\newblock {Quasinormal modes and holography}.
\newblock {\em Phys. Rev. D}, 72:086009, 2005.

\bibitem{Son:2007vk}
Dam~T. Son and Andrei~O. Starinets.
\newblock {Viscosity, Black Holes, and Quantum Field Theory}.
\newblock {\em Ann. Rev. Nucl. Part. Sci.}, 57:95--118, 2007.

\bibitem{Romatschke:2007mq}
Paul Romatschke and Ulrike Romatschke.
\newblock {Viscosity Information from Relativistic Nuclear Collisions: How
  Perfect is the Fluid Observed at RHIC?}
\newblock {\em Phys. Rev. Lett.}, 99:172301, 2007.

\bibitem{Horowitz:1999jd}
Gary~T. Horowitz and Veronika~E. Hubeny.
\newblock {Quasinormal modes of AdS black holes and the approach to thermal
  equilibrium}.
\newblock {\em Phys. Rev. D}, 62:024027, 2000.

\bibitem{Cardoso:2003cj}
Vitor Cardoso, Roman Konoplya, and Jose P.~S. Lemos.
\newblock {Quasinormal frequencies of Schwarzschild black holes in anti-de
  Sitter space-times: A Complete study on the asymptotic behavior}.
\newblock {\em Phys. Rev. D}, 68:044024, 2003.

\bibitem{Konoplya:2002zu}
R.~A. Konoplya.
\newblock {On quasinormal modes of small Schwarzschild-anti-de Sitter black
  hole}.
\newblock {\em Phys. Rev. D}, 66:044009, 2002.

\bibitem{Konoplya:2022pbc}
R.~A. Konoplya and A.~Zhidenko.
\newblock {First few overtones probe the event horizon geometry}.
\newblock {\em arXiv:2209.00679 [gr-qc],
  https://doi.org/10.48550/arXiv.2209.00679}, 9 2022.

\bibitem{Giesler:2019uxc}
Matthew Giesler, Maximiliano Isi, Mark~A. Scheel, and Saul Teukolsky.
\newblock {Black Hole Ringdown: The Importance of Overtones}.
\newblock {\em Phys. Rev. X}, 9(4):041060, 2019.

\bibitem{Konoplya:2023ahd}
R.~A. Konoplya, D.~Ovchinnikov, and B.~Ahmedov.
\newblock {Bardeen spacetime as a quantum corrected Schwarzschild black hole:
  Quasinormal modes and Hawking radiation}.
\newblock {\em Phys. Rev. D}, 108(10):104054, 2023.

\bibitem{Bolokhov:2023ruj}
S.~V. Bolokhov.
\newblock {Long lived quasinormal modes and telling oscillatory tails of the
  Bardeen spacetime}.
\newblock {\em Preprints 2023, 2023100517,
  https://doi.org/10.20944/preprints202310.0517.v1}, 2023.

\bibitem{Bolokhov:2023bwm}
S.~V. Bolokhov.
\newblock {Long-lived quasinormal modes and overtones' behavior of the holonomy
  corrected black holes}.
\newblock {\em arXiv:2311.05503 [gr-qc],
  https://doi.org/10.48550/arXiv.2311.05503}, 11 2023.

\bibitem{Konoplya:2023ppx}
R.~A. Konoplya.
\newblock {Quasinormal modes and grey-body factors of regular black holes with
  a scalar hair from the Effective Field Theory}.
\newblock {\em JCAP}, 07:001, 2023.

\bibitem{Konoplya:2023aph}
R.~A. Konoplya, Z.~Stuchlik, A.~Zhidenko, and A.~F. Zinhailo.
\newblock {Quasinormal modes of renormalization group improved Dymnikova
  regular black holes}.
\newblock {\em Phys. Rev. D}, 107(10):104050, 2023.

\bibitem{Konoplya:2022iyn}
R.~A. Konoplya.
\newblock {Quasinormal modes in higher-derivative gravity: Testing the black
  hole parametrization and sensitivity of overtones}.
\newblock {\em Phys. Rev. D}, 107(6):064039, 2023.

\bibitem{Konoplya:2022hll}
R.~A. Konoplya, A.~F. Zinhailo, J.~Kunz, Z.~Stuchlik, and A.~Zhidenko.
\newblock {Quasinormal ringing of regular black holes in asymptotically safe
  gravity: the importance of overtones}.
\newblock {\em JCAP}, 10:091, 2022.

\bibitem{Rezzolla:2014mua}
Luciano Rezzolla and Alexander Zhidenko.
\newblock {New parametrization for spherically symmetric black holes in metric
  theories of gravity}.
\newblock {\em Phys. Rev. D}, 90(8):084009, 2014.

\bibitem{Konoplya:2016jvv}
Roman Konoplya, Luciano Rezzolla, and Alexander Zhidenko.
\newblock {General parametrization of axisymmetric black holes in metric
  theories of gravity}.
\newblock {\em Phys. Rev. D}, 93(6):064015, 2016.

\bibitem{Kokkotas:2017ymc}
K.~D. Kokkotas, R.~A. Konoplya, and A.~Zhidenko.
\newblock {Analytical approximation for the Einstein-dilaton-Gauss-Bonnet black
  hole metric}.
\newblock {\em Phys. Rev. D}, 96(6):064004, 2017.

\bibitem{Kokkotas:2017zwt}
K.~Kokkotas, R.~A. Konoplya, and A.~Zhidenko.
\newblock {Non-Schwarzschild black-hole metric in four dimensional higher
  derivative gravity: analytical approximation}.
\newblock {\em Phys. Rev. D}, 96:064007, 2017.

\bibitem{Konoplya:2019fpy}
Roman~A. Konoplya, Thomas Pappas, and Alexander Zhidenko.
\newblock {Einstein-scalar\textendash{}Gauss-Bonnet black holes: Analytical
  approximation for the metric and applications to calculations of shadows}.
\newblock {\em Phys. Rev. D}, 101(4):044054, 2020.

\bibitem{Younsi:2016azx}
Ziri Younsi, Alexander Zhidenko, Luciano Rezzolla, Roman Konoplya, and Yosuke
  Mizuno.
\newblock {New method for shadow calculations: Application to parametrized
  axisymmetric black holes}.
\newblock {\em Phys. Rev. D}, 94(8):084025, 2016.

\bibitem{Konoplya:2023owh}
R.~A. Konoplya and A.~Zhidenko.
\newblock {General black-hole metric mimicking Schwarzschild spacetime}.
\newblock {\em JCAP}, 08:008, 2023.

\bibitem{Konoplya:2022tvv}
R.~A. Konoplya and A.~Zhidenko.
\newblock {Quasinormal ringing of general spherically symmetric parametrized
  black holes}.
\newblock {\em Phys. Rev. D}, 105(10):104032, 2022.

\bibitem{Abdikamalov:2021zwv}
Askar~B. Abdikamalov, Dimitry Ayzenberg, Cosimo Bambi, Sourabh Nampalliwar, and
  Ashutosh Tripathi.
\newblock {Constraining the Konoplya-Rezzolla-Zhidenko deformation parameters:
  Limits from supermassive black hole x-ray data}.
\newblock {\em Phys. Rev. D}, 104(2):024058, 2021.

\bibitem{Ni:2016uik}
Yueying Ni, Jiachen Jiang, and Cosimo Bambi.
\newblock {Testing the Kerr metric with the iron line and the KRZ
  parametrization}.
\newblock {\em JCAP}, 09:014, 2016.

\bibitem{Konoplya:2021slg}
R.~A. Konoplya and A.~Zhidenko.
\newblock {Shadows of parametrized axially symmetric black holes allowing for
  separation of variables}.
\newblock {\em Phys. Rev. D}, 103(10):104033, 2021.

\bibitem{Shashank:2021giy}
Swarnim Shashank and Cosimo Bambi.
\newblock {Constraining the Konoplya-Rezzolla-Zhidenko deformation parameters
  III: Limits from stellar-mass black holes using gravitational-wave
  observations}.
\newblock {\em Phys. Rev. D}, 105(10):104004, 2022.

\bibitem{Nampalliwar:2019iti}
Sourabh Nampalliwar, Shuo Xin, Shubham Srivastava, Askar~B. Abdikamalov,
  Dimitry Ayzenberg, Cosimo Bambi, Thomas Dauser, Javier~A. Garcia, and
  Ashutosh Tripathi.
\newblock {Testing General Relativity with X-ray reflection spectroscopy: The
  Konoplya-Rezzolla-Zhidenko parametrization}.
\newblock {\em Phys. Rev. D}, 102(12):124071, 2020.

\bibitem{Konoplya:2021qll}
R.~A. Konoplya, J.~Kunz, and A.~Zhidenko.
\newblock {Blandford-Znajek mechanism in the general stationary
  axially-symmetric black-hole spacetime}.
\newblock {\em JCAP}, 12(12):002, 2021.

\bibitem{Konoplya:2020jgt}
R.~A. Konoplya, A.~F. Zinhailo, and Z.~Stuchlik.
\newblock {Quasinormal modes and Hawking radiation of black holes in cubic
  gravity}.
\newblock {\em Phys. Rev. D}, 102(4):044023, 2020.

\bibitem{Konoplya:2020hyk}
R.~A. Konoplya and A.~Zhidenko.
\newblock {General parametrization of black holes: The only parameters that
  matter}.
\newblock {\em Phys. Rev. D}, 101(12):124004, 2020.

\bibitem{Konoplya:2022kld}
R.~A. Konoplya and A.~Zhidenko.
\newblock {How general is the strong cosmic censorship bound for quasinormal
  modes?}
\newblock {\em JCAP}, 11:028, 2022.

\bibitem{Konoplya:2020kqb}
Roman~A. Konoplya, Thomas~D. Pappas, and Zdeněk Stuchlík.
\newblock {General parametrization of higher-dimensional black holes and its
  application to Einstein-Lovelock theory}.
\newblock {\em Phys. Rev. D}, 102(8):084043, 2020.

\bibitem{Mannheim:1988dj}
Philip~D. Mannheim and Demosthenes Kazanas.
\newblock {Exact Vacuum Solution to Conformal Weyl Gravity and Galactic
  Rotation Curves}.
\newblock {\em Astrophys. J.}, 342:635--638, 1989.

\bibitem{Boulware:1985wk}
David~G. Boulware and Stanley Deser.
\newblock {String Generated Gravity Models}.
\newblock {\em Phys. Rev. Lett.}, 55:2656, 1985.

\bibitem{Konoplya:2017zwo}
R.~A. Konoplya and A.~Zhidenko.
\newblock {Quasinormal modes of Gauss-Bonnet-AdS black holes: towards
  holographic description of finite coupling}.
\newblock {\em JHEP}, 09:139, 2017.

\bibitem{Fortuna:2020obg}
Sean Fortuna and Ian Vega.
\newblock {Bernstein spectral method for quasinormal modes and other eigenvalue
  problems}.
\newblock {\em Eur. Phys. J. C}, 83(12):1170, 2023.

\bibitem{Konoplya:2022xid}
R.~A. Konoplya and A.~Zhidenko.
\newblock {Nonoscillatory gravitational quasinormal modes and telling tails for
  Schwarzschild\textendash{}de Sitter black holes}.
\newblock {\em Phys. Rev. D}, 106(12):124004, 2022.

\bibitem{Konoplya:2022zav}
R.~A. Konoplya and A.~Zhidenko.
\newblock {Bernstein spectral method for quasinormal modes of a generic black
  hole spacetime and application to instability of dilaton\textendash{}de
  Sitter solution}.
\newblock {\em Phys. Rev. D}, 107(4):044009, 2023.

\bibitem{Konoplya:2017ymp}
R.~A. Konoplya and A.~Zhidenko.
\newblock {Eikonal instability of
  Gauss-Bonnet\textendash{}(anti-)\textendash{}de Sitter black holes}.
\newblock {\em Phys. Rev. D}, 95(10):104005, 2017.

\bibitem{Natario:2004jd}
Jose Natario and Ricardo Schiappa.
\newblock {On the classification of asymptotic quasinormal frequencies for
  d-dimensional black holes and quantum gravity}.
\newblock {\em Adv. Theor. Math. Phys.}, 8(6):1001--1131, 2004.

\bibitem{Arean:2023ejh}
Daniel Areán, David~García Fariña, and Karl Landsteiner.
\newblock {Pseudospectra of holographic quasinormal modes}.
\newblock {\em JHEP}, 12:187, 2023.

\bibitem{Gonzalez:2017gwa}
P.~A. González, R.~A. Konoplya, and Yerko Vásquez.
\newblock {Quasinormal modes of a scalar field in the Einstein-Gauss-Bonnet-AdS
  black hole background: Perturbative and nonperturbative branches}.
\newblock {\em Phys. Rev. D}, 95(12):124012, 2017.

\bibitem{Dotti:2005sq}
Gustavo Dotti and Reinaldo~J. Gleiser.
\newblock {Linear stability of Einstein-Gauss-Bonnet static spacetimes. Part I.
  Tensor perturbations}.
\newblock {\em Phys. Rev. D}, 72:044018, 2005.

\bibitem{Gleiser:2005ra}
Reinaldo~J. Gleiser and Gustavo Dotti.
\newblock {Linear stability of Einstein-Gauss-Bonnet static spacetimes. Part
  II: Vector and scalar perturbations}.
\newblock {\em Phys. Rev. D}, 72:124002, 2005.

\bibitem{Grozdanov:2016vgg}
Sašo Grozdanov, Nikolaos Kaplis, and Andrei~O. Starinets.
\newblock {From strong to weak coupling in holographic models of
  thermalization}.
\newblock {\em JHEP}, 07:151, 2016.

\bibitem{Cheung:2021bol}
Mark Ho-Yeuk Cheung, Kyriakos Destounis, Rodrigo~Panosso Macedo, Emanuele
  Berti, and Vitor Cardoso.
\newblock {Destabilizing the Fundamental Mode of Black Holes: The Elephant and
  the Flea}.
\newblock {\em Phys. Rev. Lett.}, 128(11):111103, 2022.

\bibitem{Berti:2022xfj}
Emanuele Berti, Vitor Cardoso, Mark Ho-Yeuk Cheung, Francesco Di~Filippo,
  Francisco Duque, Paul Martens, and Shinji Mukohyama.
\newblock {Stability of the fundamental quasinormal mode in time-domain
  observations against small perturbations}.
\newblock {\em Phys. Rev. D}, 106(8):084011, 2022.

\bibitem{Konoplya:2023hqb}
R.~A. Konoplya.
\newblock {The sound of the event horizon}.
\newblock {\em Int. J. Mod. Phys. D}, 32(14):2342014, 2023.

\bibitem{Takahashi:2010ye}
Tomohiro Takahashi and Jiro Soda.
\newblock {Master Equations for Gravitational Perturbations of Static Lovelock
  Black Holes in Higher Dimensions}.
\newblock {\em Prog. Theor. Phys.}, 124:911--924, 2010.

\bibitem{Cuyubamba:2016cug}
M.~A. Cuyubamba, R.~A. Konoplya, and A.~Zhidenko.
\newblock {Quasinormal modes and a new instability of Einstein-Gauss-Bonnet
  black holes in the de Sitter world}.
\newblock {\em Phys. Rev. D}, 93(10):104053, 2016.

\end{thebibliography}

\end{document}